# Plasmodium Detection Using Simple CNN and Clustered GLCM Features

Julisa Bana Abraham
Electrical and Information Engineering
Universitas Gadjah Mada
Yogyakarta, Indonesia
julisa.bana.abraham@mail.ugm.ac.id

*Abstract*— Malaria is a serious disease caused by the Plasmodium parasite that transmitted through the bite of a female Anopheles mosquito and invades human erythrocytes. Malaria must be recognized precisely in order to treat the patient in time and to prevent further spread of infection. The standard diagnostic technique using microscopic examination is inefficient, the quality of the diagnosis depends on the quality of blood smears and experience of microscopists in classifying and counting infected and non-infected cells. Convolutional Neural Networks (CNN) is one of deep learning class that able to automate feature engineering and learn effective features that could be very effective in diagnosing malaria. This study proposes an intelligent system based on simple CNN for detecting malaria parasites through images of thin blood smears. The CNN model obtained high sensitivity of 97% and relatively high PPV of 81%. This study also proposes a false positive reduction method using feature clustering extracted from the gray level co-occurrence matrix (GLCM) from the Region of Interests (ROIs). Adding the GLCM feature can significantly reduce false positives. However, this technique requires manual set up of silhouette and euclidean distance limits to ensure cluster quality, so it does not adversely affect sensitivity.

## I. INTRODUCTION

Malaria is an infectious disease caused by the Plasmodium parasite that transmitted through the bite of a female Anopheles mosquito and invades human erythrocytes. Malaria must be recognized precisely in order to treat the patient in time and to prevent further spread of infection. Various types of parasites include P. ovale, P. malariae, P. vivax, and P. falciparum can infect humans. P. vivax causes the most of malaria cases. However, P. falciparum induces the most serious and sometimes fatal type of malaria. In 2016, the World Health Organization (WHO) reported 212 million cases of the disease worldwide [1]. Malaria parasites can be identified by microscopic examination of thick and thin blood smears. Thick blood smears are used to detect the existence of parasites while thin blood smears assist in identifying species of parasites that cause infection [2]. However, the accuracy is very dependent on the quality of the smear and experience of microscopists in classifying and counting infected and non-infected cells.

Other than microscopic examination, there are several methods that can be used for malaria detection. Some advancements have been made in leveraging image processing techniques to extract hand-engineered features and build machine learning-based classification models. However it takes a lot of time and still requires expertise in analyzing morphological variations, textures, and positions of the region of interest (ROI) in the figure. For images, an important source of information is in the local correlation between an adjacent pixel. Convolutional Neural Networks (CNN), is one class in the DL model that able to learn different aspects of images. It allows CNNs to automate feature engineering and learn effective features that generalize well on new data points. Accurate diagnosis is critical to the effective management of malaria, CNN model promises highly scalable results and excels in the extraction and classification of end-to-end features that could be very effective in diagnosing malaria.

## II. RELATED WORKS

Convolutional Neural Networks for detection of malaria parasites has been used in several previous studies. Rajaraman et al [3] evaluated the performance of deep learning-based pre-trained CNN models as feature extractor in classifying infected and non-infected cells. This study used the Laplacian of Gaussian (LoG) to obtain the region of interest (ROI) and cross-validates the performance of predictive models at the patient level to reduce bias and errors in generalization. The results of segmentation and detection based on the proposed segmentation model show a positive predictive value (PPV) of 94.44% with a sensitivity value of 96.20%. Based on the results of the study it can be concluded that pre-trained ResNet-50 [4] is superior in classifying infected and non-infected cells.

Peñas et al [5] built a system that can detect the presence of malaria parasites and identify the type of parasite species in blood samples. In this study, a series of morphological transformations were implemented for pre-processing before segmentation using Connected Components Analysis, and CNN for classification of data. The results of the study showed an accuracy of 92.40% for parasite detection and 87.90% for identification of parasite types.

Liang et al. [6] used the deep learning approach to detect parasitic blood cells, this study designed a new CNN model for classification of blood cells. The results show that the proposed CNN model, that has 17 layers, has superior performance compared to the transfer learning model using pre-trained AlexNet [7] as the feature extractor. The average classification accuracy of the CNN model is 97.37%, with the sensitivity, specificity, and precision of the models all reaching the level of 97%.

From the review that has been done, it shows that CNN has good performance in detecting and classifying *Plasmodium* images, but many of the previous studies only perform the classification on the cropped thin blood smear image which is actually not practical in the real world.

This paper proposes a method for detecting *Plasmodium* in thin blood smear images based on the CNN model. In this paper, we compare the performance of two CNN architectures: simple CNN with two convolution layers and transfer learning model with pre-trained ResNet-50 as the feature extractor.

## III. METHODS

This paper proposes a *Plasmodium* detection method based on CNN from the uncropped thin blood smear images and provides a bounding-box that matches the size of *Plasmodium*.

### A. Dataset

There are two datasets used for this study: training and testing datasets. The training dataset contains cropped thin blood smear image that was divided into two classes: *Plasmodium* and non-*Plasmodium*, which amounts to 330. The number of images in each class is unbalanced, the *Plasmodium* class contains 148 images and non-Plasmodium contains 182 images.

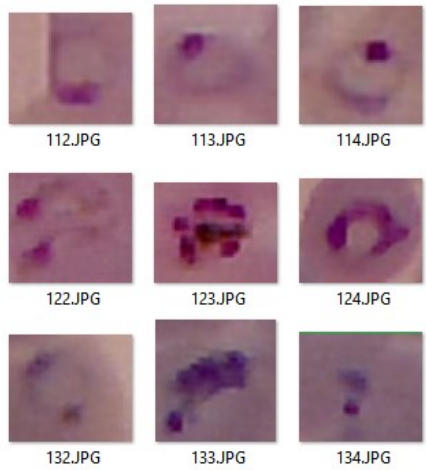

Figure 1. Training dataset contains cropped thin blood smear image

The training dataset were obtained from cropped ground truth from the testing images. While testing dataset contains 35 images of uncropped thin blood smear that have dimensions of 1280x960 pixels in RGB format.

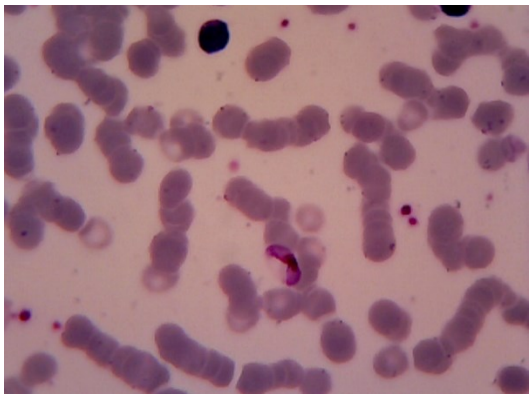

Figure 2. Testing dataset contains uncropped thin blood smear image

### B. Training Data Preprocessing

The training images were read in Python 3.7 with OpenCV module and each of the training image is then resized to 100x100 pixels and divided by 255 so that the value on each RGB channel is in the range 0-1.

### C. Data Augmentation

The training dataset that contains 330 pieces of images from thin blood smear has two classes, namely *Plasmodium* and non-*Plasmodium* was considered too small and imbalanced to train the CNN model. Data augmentation is a technique for adding CNN training data by translating, rotating and zooming with certain values on each training image. We use data augmentation to avoid local optimum or widely called overfitting during the training process of the CNN model.

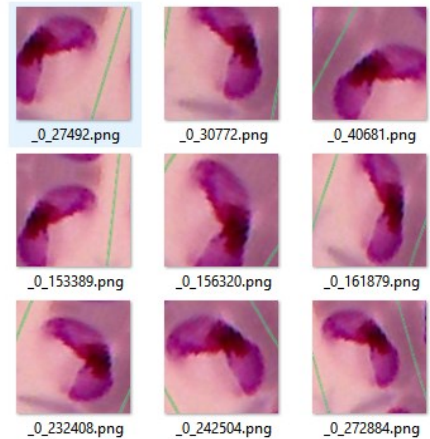

Figure 3. Data augmentation ilustration

### D. CNN Model

This paper uses keras 2.2.4 module in Python 3.7 programming language for the CNN modeling. The CNN model that we propose uses two convolutional layers and two fully connected layers. Our model also adds 2 drop out layers to reduce overfitting with a probability of 0.5 each. LeNet [6] inspires the architecture of this model. Our CNN architecture is shown in Figure 4.

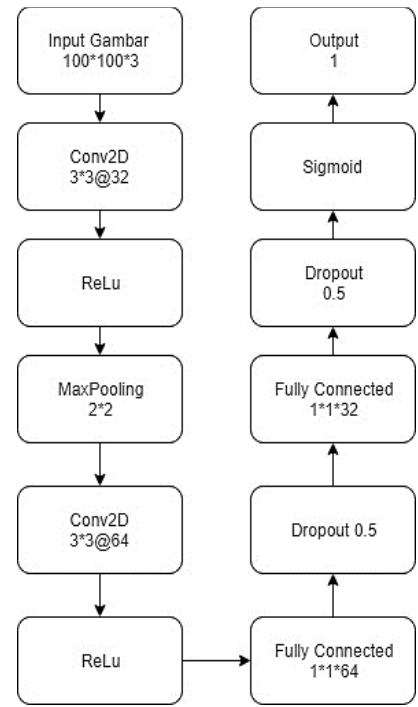

Figure 4. Proposed CNN model

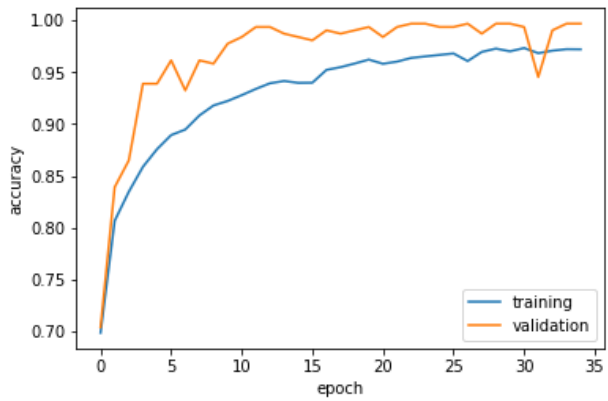

Figure 5. Epoch vs accuracy from training process

The CNN model training obtained 97% training accuracy and 99% validation accuracy. Validation data were obtained from 100 random images obtained from the data augmentation process, so there is a possibility that the high validation accuracy was due to the overfitting. The graph of the training process can be seen in Figure 5.

*E. Detection Process*

In this paper, ROI is obtained from a process based on the Laplacian of Gaussian (LoG) [7] inspired by the work of Rajaraman et al [3]. ROI generation using LoG requires less time and more robust than other methods such as the sliding window. The detection process in the testing dataset is shown in Figure 6.

*1) HSV Colorspace Conversion*

In this paper, we use saturation channels on HSV colorspace to obtain the greyscale image. Because the saturation of the *Plasmodium* tends to be higher than the red blood cell (RBC) and background, so that it can easily be spotted.

*2) Multi-Scale Laplacian of Gaussian (LoG)*

Laplacian of Gaussian is one of the most commonly used methods for blob detection. Blob is a bright part of a dark region or vice versa. The location of the blob obtained from multi-scale LoG performed on the grayscale saturation image coincides with the location of *Plasmodium*, so that we can apply the classifier to the location of the blob to determine the presence of *Plasmodium* [3]. We performed a multi-scale LoG with stages between 4-7 on the grayscale saturation image to ensure all blobs can be detected.

*3) Otsu Thresholding*

Otsu thresholding [8] is performed on multi-scale LoG image that indicates the location of blobs. Otsu thresholding is chosen because it is fast and returns a good binary image so that makes it easy to find the center of the contour to be used as ROI.

*4) Opening*

Opening is a logical operation that combines the erosion and dilation process. Opening is the dilation of the erosion of a set A by a structuring element B. The opening is done to clean the contour from noise.

*5) Dilation*

Dilation is done to enlarge an object in a binary image. In this paper, this process is done to improve the accuracy of contour centroids.

*6) Grey Level Co-occurrence Matrix Clustered features*

To reduce the number of false positives, we use Grey Level Co-occurrence Matrix (GLCM) [9] to extract textural features from all ROI detected by the CNN model. The image-0.15.0 module can be used to obtain five statistical values that can be obtained from GLCM, namely 'dissimilarity', 'correlation' contrast, 'homogeneity', and 'energy'. However, in this paper we only use the value of 'homogeneity', and 'energy'. The reason we only use the value of 'homogeneity', and 'energy' for the clustering process is that those values can be linearly separated visually. The results of clustering on one of the thin blood smear images ROIs are shown in Figure 7.

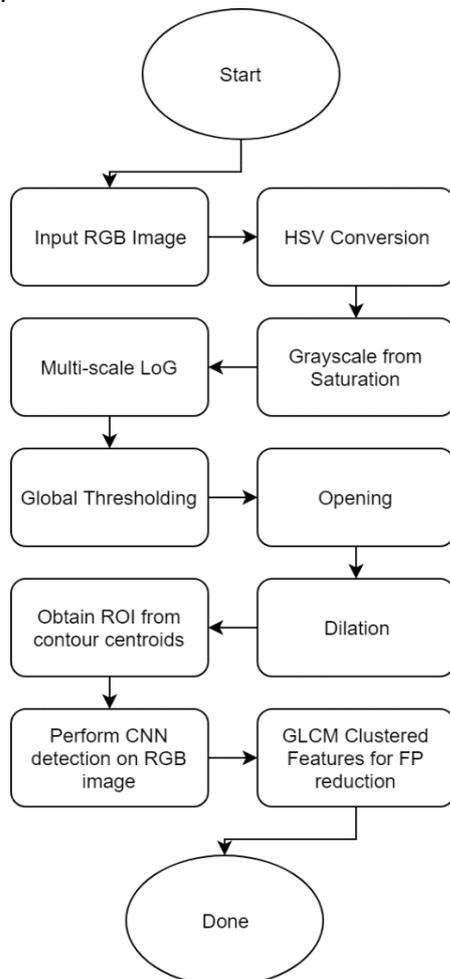

Figure 6. ROI generation flowchart

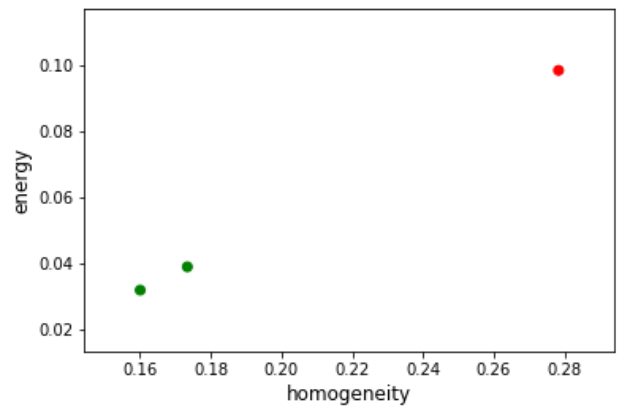

Figure 7. GLCM extracted homogeneity and energy

After extraction of textural features using GLCM, clustering is done to be able to separate the *Plasmodium* from the other false positives. We use K-Means Clustering because the class to be separated is already known, including plasmodium, white blood cells, and red blood cells. However, in an uncropped thin blood smear image maybe the three classes did not appear entirely so to determine the value of k we use silhouette metrics, by choosing k that has the highest silhouette value. As shown in Figure 8 based on the results of the silhouette, it was chosen to use k = 2 to cluster the entire class well. From our observation, the cluster containing *Plasmodium* usually have the lowest homogeneity and energy among other clusters.

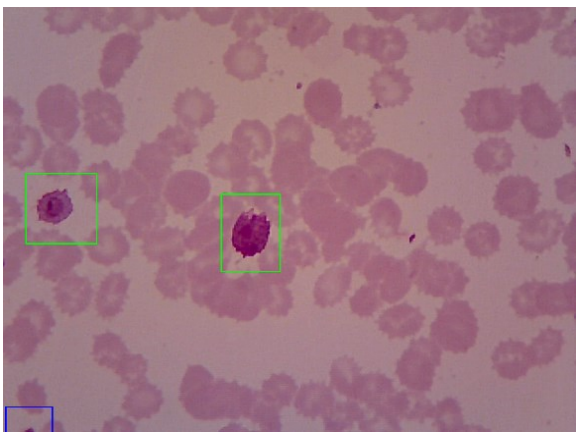

Figure 8. Clustering result of the GLCM features, the green and blue box indicate true positive and false positive respectively

## IV. RESULT AND DISCUSSION

An illustration of the ROI generation step on one of the testing images is shown in Figure 9.

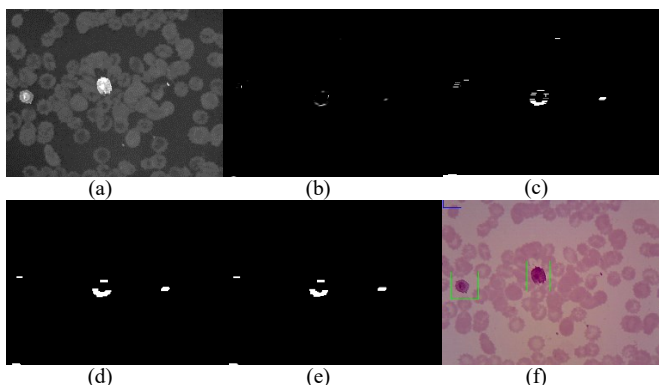

Figure 9. Image processing illustration: (a) saturation grayscale, (b) LoG, (c) Otsu Thresholding, (d) Opening, (e) Dilation, (f) Bounding-boxed image

The evaluation to determine the performance of the model uses a comparison between the bounding box on ground truth and the detection bounding box. If the detection bounding box is in the same region as the bounding box on the ground truth the detection results are true positive (TF). If there is a detection bounding box but not at the ROI location on the ground truth it is a False Positive (FP), if there is no detection bounding box but there is a ground truth bounding box then it is a false negative (FN).

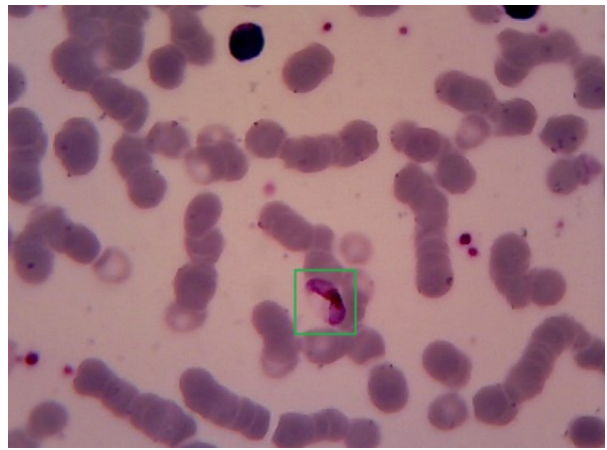

Figure 10. Testing image containing ground truth

The performance metrics used are positive predictive value (PPV) which can be seen in equation 1 and the sensitivity can be seen in equation 2.

$$PPV = \frac{TP}{TP + FP} \times 100\% \quad (1)$$

$$Sensitivity = \frac{TP}{TP + FN} \times 100\% \quad (2)$$

The evaluation was carried out with positive predictive value (PPV) and sensitivity. Confusion matrix from the evaluation is shown in Table 1

Table 1. Confusion Matrix

|        |     | Predicted |         |
|--------|-----|-----------|---------|
|        |     | No        | Yes     |
| Actual | No  | TN = NaN  | FP = 15 |
|        | Yes | FN = 2    | TP = 64 |

From the confusion matrix, we can calculate the PPV of 81% and sensitivity of 97%. The evaluation results can be said to be very good even though it only uses simple CNN that only has two convolution layers that proven enough to extract features from the cropped thin blood smear image. The implementation of drop out layer is also proven can reduce overfitting as evidenced by the high results of the validation in the training process and yield high sensitivity on testing dataset that contains uncropped thin blood smear images.

These results show our model produces relatively few false positives. The low False Positive is partly caused by GLCM clustered features, which can separate between clusters containing Plasmodium with false positive. However, the determination of the limit of silhouette score greatly affects the results of PPV and sensitivity because if the value of the silhouette is large enough, which means that clustering, is good but it can separate the true positives so that it can reduce the value of sensitivity. Therefore, it is necessary to apply the silhouette boundary value to ensure if the clustering is does not separate true positives. In this study, the boundary value of the silhouette and euclidean distance which if only consists of two ROI is done manually for each testing image. However, the typical ranges from 0.45-0.55 for silhouette and 0.05-0.07 for the euclidean distance.

The size of the bounding-box for ROI also affects the accuracy of CNN detection, so in this study, besides using a combination of multi-stage LoG we also combined several bounding-box sizes. This technique successfully increases the sensitivity of our detector. The effect of the bounding-box size may be due to our CNN model being overfitting due to little data training so that size differences also affect the accuracy of detection, but this only occurs in some images.

CNN is considered as a blackbox method because we do not know how CNN works internally, but at least we can visualize the intermediate activation layer that can be done using Keras [10]. Intermediate activation is useful for understanding how the convolution layer transforms its inputs and gets the initial concepts of individual CNN filters. The visualization of the intermediate activation layer for our trained model is shown in Figure 11.

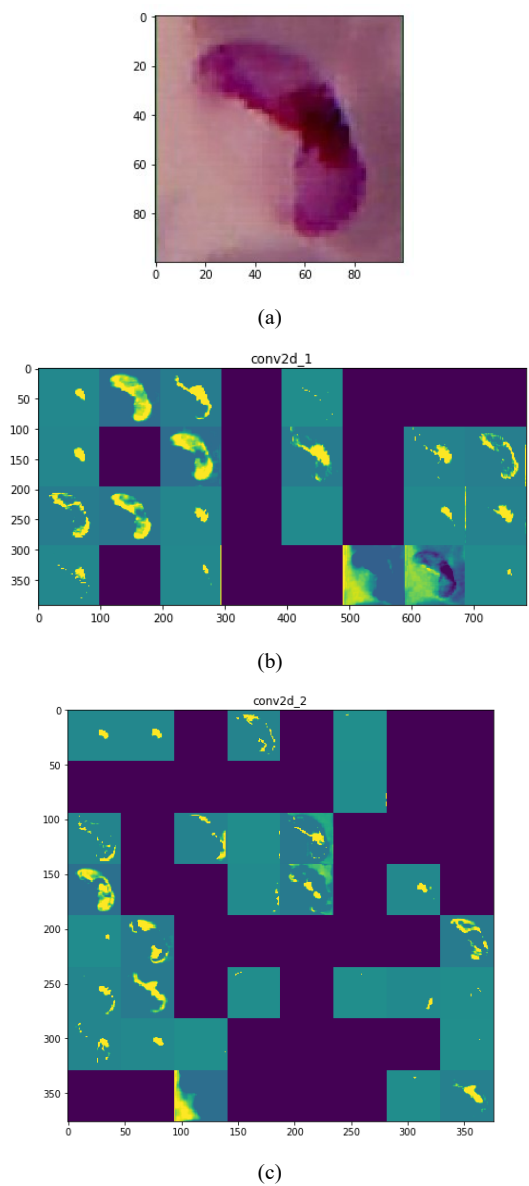

Figure 11. How CNN sees the *Plasmodium*: (a) input image, (b) first convolution layer activation (c) second convolution layer activation

As can be seen in Figure 12, actually CNN has encoded the segmentation, horizontal and vertical edge detection and several specific morphology detector of the *Plasmodium* automatically. The blank spot means that there is no pattern on the input image encode by the specific filter. Then after that, the CNN model performs the training process using the fully connected layer to be able to build the classifier using the features extracted by the convolution layer.

The detection scheme that we propose has a computation time of around 2-3 seconds per image on the Intel Core i7 4700MQ and Nvidia GT750M which is the middle-level CPU and GPU at this time. In our opinion, the computation time is fast enough hence it can be applied to the real-world problems, but this requires the re-training of the CNN model to adjust the blood smear characteristics.

V. CONCLUSION

From the performance metrics evaluation, we can conclude that simple CNN trained with augmented small training data has been able to accurately detect *Plasmodium* in the uncropped thin blood smear data, as evidenced by good sensitivity of 97% and with relatively high PPV of 81%.

The GLCM extracted features that we use in this study help to reduce the false positives generated. However, the need for manually set the silhouette and euclidean distance limits to ensure cluster quality so that it does not adversely affect sensitivity is a weakness of this technique.